 \documentclass[final,5p,times,twocolumn,authoryear]{elsarticle}

\usepackage{amssymb}
\usepackage{lipsum}
\usepackage{hyperref}
\usepackage{graphicx}

\journal{Astronomy $\&$ Computing}

\begin{document}

\begin{frontmatter}



\title{The Array Control and Data Acquisition software of the Cherenkov Telescope Array Observatory }


\author[CIEMAT]{I.~Oya\corref{cor1}}
\ead{igor.oya@ciemat.es}

\author[CTAO]{B.~López\corref{cor1}}
\ead{bernhard.lopez@cta-observatory.org}
\cortext[cor1]{Corresponding Author}


\author[LAPP]{P.~Aubert}
\author[UniGeneva]{G.~Barni}
\author[IEEC]{P.~Bauza}
\author[DESY]{D.~Berge}
\author[DESY]{J.-P.~Bolle}
\author[CTAO]{W.~Boulakbech}
\author[INAF-Catania]{P.~Bruno}
\author[COSYLAB]{U.~Bajc}
\author[INAF-Bologna]{A.~Bulgarelli}
\author[INAF-Bologna]{M.~Cappi}
\author[CPPM]{F.~Cassol}
\author[LAPP]{S.~Caroff}
\author[INAF-Bologna]{L.~Castaldini}
\author[UniPotsdam]{T.~Collins}
\author[INAF-Bologna]{V.~Conforti}
\author[INAF-Catania]{A.~Costa}
\author[INRIA]{L.~David}
\author[INAF-Bologna]{G.~De Cesare}
\author[DESY]{E.~de~Ona~Wilhelmi}
\author[INAF-Bologna]{A.~Di~Piano}
\author[UniPotsdam]{K.~Egberts}
\author[COSYLAB]{R.~Fernandez} 
\author[INAF-Bologna]{V.~Fioretti}
\author[UniGeneva]{S.~Fukami}
\author[ICE-CSIC,IEEC]{E.~García}
\author[LAPP]{E.~Garcia}
\author[DESY]{H.~Gasparyan} 
\author[UniPerugia]{S.~Germani}
\author[MPIK]{J.~Hinton}
\author[UniPotsdam]{C.~Hoischen}
\author[INAF-Catania]{F.~Incardona}
\author[DESY]{D.~Kostunin}
\author[UniGeneva]{E.~Lyard}
\author[LAPP]{G.~Maurin}
\author[DESY]{D.~Melkumyan}
\author[ICE-CSIC]{E.~Mestre}
\author[INAF-Catania]{K.~Munari}
\author[DESY]{T.~Murach}
\author[CAMK]{A.~Muraczewski}
\author[ICE-CSIC]{N.~Nakhjiri}
\author[CTAO]{D.~Neise}
\author[LAPP]{T.~Oprinsen}
\author[INAF-Bologna]{G.~Panebianco} 
\author[INAF-Bologna]{N.~Parmiggiani}
\author[INRIA]{E.~Pietriga}
\author[LAPP]{V.~Pollet}
\author[CAMK]{B. Rudak}
\author[DESY]{I.~Sadeh}
\author[COSYLAB]{S.~Sah}
\author[DESY]{A.~Sarkar}
\author[UniGeneva]{M.~Schefer}
\author[DESY]{T.~Schmidt}
\author[ICE-CSIC,IEEC]{D.~Soldevila}
\author[INAF-Catania]{S.~Spinello}
\author[CTAO]{C.~Steppa}
\author[ICE-CSIC,IEEC,ICREA]{D.~F.~Torres} 
\author[UniGeneva]{A.~Tramacere}
\author[ICE-CSIC,IEEC]{R.~Vallés} 
\author[LAPP]{T.~Vuillaume} 
\author[UniGeneva]{R.~Walter}
\author[MPIK]{F.~Werner}
\author[CTAO]{A.~Wörheide}

\author[CTAO-Web]{for~the~CTAO}

\affiliation[CIEMAT]{organization={CIEMAT},
            addressline={Avda. Complutense 40}, 
            city={Madrid},
            postcode={28040}, 
            country={Spain}}
\affiliation[CTAO]{organization={CTAO ERIC},
            addressline={Science Data Management Centre (SDMC), Platanenallee 6}, 
            city={Zeuthen},
            postcode={15738}, 
            country={Germany}}
\affiliation[LAPP]{organization={Univ. Savoie Mont Blanc, CNRS, Laboratoire d’Annecy de Physiquedes Particules-IN2P3},
            city={Annecy},
            postcode={74000}, 
            country={France}}
\affiliation[UniGeneva]{organization={Département d’Astronomie, Université de Genève},
            addressline={Chemin d’Ecogia 16}, 
            city={Versoix},
            postcode={1290}, 
            country={Switzerland}}
\affiliation[IEEC]{organization={Institut d’Estudis Espacials de Catalunya(IEEC)},
            addressline={Parc UPC - PMT, Edif. RDIT, Carrer d'Esteve Terradas, 1}, 
            city={Castelldefels},
            postcode={08860}, 
            country={Spain}}
\affiliation[DESY]{organization={Deutsches Elektronen-Synchrotron DESY},
            addressline={Platanenallee 6}, 
            city={Zeuthen},
            postcode={15738}, 
            country={Germany}}
\affiliation[INAF-Catania]{organization={INAF, Osservatorio Astrofisico di Catania},
            addressline={Via S. Sofia 78}, 
            city={Catania},
            postcode={95123}, 
            country={Italy}}
\affiliation[COSYLAB]{organization={COSYLAB JSC, Control System Laboratory},
            addressline={Gerbičeva ulica 64}, 
            city={Ljubljana},
            postcode={1000}, 
            country={Slovenia}} 
\affiliation[INAF-Bologna]{organization={INAF/OAS Bologna},
            addressline={Via Gobetti 93/3}, 
            city={Bologna},
            postcode={40129}, 
            country={Italy}}
\affiliation[CPPM]{organization={CNRS/IN2P3, CPPM},
            addressline={Aix Marseille Univ}, 
            city={Marseille},
            postcode={13288}, 
            country={France}} 
\affiliation[UniPotsdam]{organization={Institut für Physik \& Astronomie, Universität Potsdam},
            addressline={Karl-Liebknecht-Strasse 24/25}, 
            city={Potsdam},
            postcode={14476}, 
            country={Germany}}
\affiliation[INRIA]{organization={Université Paris-Saclay, CNRS, Inria},
            city={Gif-sur-Yvette},
            country={France}}
\affiliation[ICE-CSIC]{organization={Institute of Space Sciences(ICE,CSIC)},
            addressline={Carrer de Can Magrans, s/n}, 
            city={Cerdanyola del Vallés},
            postcode={08193}, 
            country={Spain}} 
\affiliation[UniPerugia]{organization={Universitá di Perugia, Dipartimento di Fisica e Geologia},
            addressline={Via Alessandro Pascoli}, 
            city={Perugia},
            postcode={06123}, 
            country={Italy}} 
\affiliation[MPIK]{organization={Max-Planck-Institut für Kernphysik},
            addressline={Saupfercheckweg 1}, 
            city={Heidelberg},
            postcode={69117}, 
            country={Germany}} 
\affiliation[CAMK]{organization={Nicolaus Copernicus Astronomical Center, Polish Academy of Sciences},
            addressline={ul. Bartycka 18}, 
            city={Warsaw},
            postcode={00-716}, 
            country={Poland}}  
\affiliation[ICREA]{organization={Institució Catalana de Recerca i Estudis Avançats (ICREA)},
            addressline={Carrer de Can Magrans, s/n}, 
            city={Cerdanyola del Vallés},
            postcode={08193}, 
            country={Spain}} 
\affiliation[CTAO-Web]{organization={https://www.ctao.org/},
           } 


\begin{abstract}
The Cherenkov Telescope Array Observatory (CTAO) aims to advance knowledge of the gamma-ray sky as the largest gamma-ray observatory ever built. The CTAO will be deployed at two sites, one in the Northern Hemisphere and the other in the Southern Hemisphere, containing telescopes of three sizes to cover different energy domains. Commissioning of the prototype CTAO Large-Sized Telescope (LST-1) is being finalized at the northern site, while three additional LSTs are under construction. Additional calibration and environmental monitoring instruments, such as laser imaging detection and ranging (LIDAR) systems and weather stations, will support telescope operations. The Array Control and Data Acquisition (ACADA) system serves as the central element for on-site CTAO operations. ACADA controls, supervises, and handles the data generated by the telescopes and the auxiliary instruments. It drives the efficient planning and execution of observations while managing the multi-gigabit-per-second data streams produced by each CTAO telescope. The ACADA system contains the CTAO Science Alert Generation Pipeline – a real-time data processing and analysis pipeline, dedicated to automatically generating science alert candidates as data are acquired. These science alerts, along with external alerts received from other scientific instruments, are managed by the Transients Handler (TH) component. The TH informs ACADA’s Short-Term Scheduler (STS) about relevant science alerts, enabling modification of ongoing observations on sub-minute timescales. This capability for rapid response, combined with the fast slewing of CTAO telescopes, makes the Observatory an excellent instrument for studying high-impact astronomical transients. The ACADA software is built on the Alma Common Software (ACS) framework and implemented in C++, Java, Python, and JavaScript. The first major release of the ACADA software, ACADA REL1, was completed in July 2023 and integrated following a testing campaign with the LST-1 concluded in October 2023. Since then, ACADA has passed the Critical Design Review (CDR) stage and is now being prepared for the next major release, ACADA REL2. This article describes the design and current development status of the ACADA software system.
\end{abstract}




\end{frontmatter}




\section{Introduction}
\label{introduction}

The Cherenkov Telescope Array Observatory (CTAO) will be the largest and most advanced ground-based gamma-ray observatory \citep{Science-CTA, CTAO-Stuart}. The CTAO has two telescope array sites: one in the Southern Hemisphere (CTAO-South), near the Paranal Observatory in Chile, and the other in the Northern Hemisphere (CTAO-North), at the Roque de los Muchachos Observatory (ORM) in La Palma, Spain. In total, more than 60 imaging atmospheric Cherenkov telescopes (IACTs) will be deployed across the two sites, along with additional auxiliary instruments, making the CTAO one of the world's largest astronomical installations currently under development~\citep{CTAO-Stuart}.

The Cherenkov telescopes and auxiliary instruments—collectively known as array elements (AEs)—will be controlled and supervised by operators via the Array Control and Data Acquisition (ACADA) system from control rooms at each site. ACADA is also responsible for handling, compressing, and storing the acquired data on disk. ACADA includes a pipeline for the rapid assessment of acquired data, enabling early identification of transient phenomena or data issues—the Science Alert Generation Pipeline (SAG; see Sec. \ref{sec:SAG}). 

Data acquired by ACADA are archived and processed by the Data Processing and Preservation System (DPPS), hosted in two on-site data centers adjacent to the telescopes—where ACADA also operates—and in four off-site data centers located in Europe. The Science User Support System (SUSS), managed from the CTAO Science Data Management Centre (SDMC) in Zeuthen, Germany, will make processed data and science tools available to end users. SUSS will also provide the interface for scientists to create and submit CTAO observation proposals, and will generate the long- and mid-term observation schedules. In addition, the SUSS team will provide the Science Analysis Tools (SAT) to researchers using CTAO data. The SAT software is based on the open-source Python package Gammapy \citep{gammapy:2023}. 

The ACADA collaboration was established between 2019 and 2020; however, most partners in the ACADA collaboration had been prototyping and collaborating as part of the CTAO Consortium activities during previous years~\citep{OES_ACAT, ACADA_ICALEPCS}. As a result of this groundwork, the requirements, architecture, design, and development process of ACADA were rapidly formalized, followed by successful Preliminary Design Review (PDR) and Critical Design Review (CDR) stages. Two major software releases have been completed, followed by a successful Integration and Test (I\&T) campaign with the Large-Sized Telescope Prototype (LST-1)~\citep{LST1}. With the design and development process of ACADA validated through these reviews and campaigns, the ACADA team is now developing subsequent versions to support the Observatory’s construction schedule, array-element integration, science verification, and, eventually, early science operations. 

This paper describes the design, current status, and development process of the ACADA software system.
Section.~\ref{requirements} introduces the driving requirements of ACADA, Section~\ref{architecture} presents an overview of its architecture and technology stack, and Section~\ref{sec:devel-process} describes the development process. Section~\ref{sec:status} details the ongoing development of ACADA Release 2 (REL-2), the main technical challenges, and the planned scope for future versions. Finally, Section~\ref{summary} summarizes the work.

\section{Requirements for the ACADA System}
\label{requirements}

ACADA is a central system in CTAO operations, and as such, its product requirements (known as level-B requirements in CTAO terminology) were among the first to be developed. The ACADA requirements were defined through a process that leveraged experience from previous IACT experiments, prototyping efforts, and a top-down Model-Based Systems Engineering (MBSE) approach. Nearly 300 ACADA system-level (level-B) requirements were identified, specified, and validated through a joint effort involving ACADA experts and the CTAO Systems Engineering, Science Operations, and Computing Department teams. These requirements were further validated through formal reviews, including the PDR (2020), and the CDR (2024–2025), both involving international experts experienced in building systems similar to ACADA. 

Level-B requirements were subsequently decomposed into over 1,000 subsystem-level (level-C) requirements, each allocated to one of the 11 ACADA subsystems and linked to corresponding ACADA functions, use cases (see later), components, and interfaces, following an MBSE approach \citep{ACADA-Architecture}. The following list summarizes key ACADA capabilities as defined by these requirements.

\begin{itemize}
    \item Operation of at least eight independently configurable groups of telescopes (subarrays) simultaneously, dynamically formed from distinct subsets of available telescopes at a given site.
    \item Support up to a hundred telescopes, along with light detection and ranging (LIDAR) instruments, stellar photometers, all-sky cameras, weather stations, ceilometers, and external light sources for calibrating Cherenkov telescope cameras (known as illuminators in CTAO terminology). 
    \item Collect event data from all CTAO telescope cameras, handling data volumes of approximately 24 Gb/s for each Large-Sized Telescopes (LSTs; up to four per site), 12 Gb/s for each Medium-Sized Telescopes (MSTs; up to 25 in CTAO-South and 19 in CTAO-North), and 2 Gb/s for each Small-Sized Telescopes (SSTs; up to 70 in CTAO-South). 
    \item Perform real-time data volume reduction for Cherenkov telescope cameras so that the total data rate delivered to the DPPS at each site remains below 5 Gb/s, minimizing impact on scientific performance (a reduction factor of up to 50 once the two full telescope arrays are deployed).
    \item Acquire all hardware-related monitoring points provided by the AEs, with an overall capacity of up to 200,000 data points.
    \item Handle ACADA and AE errors and alarms, providing CTAO operators with suggested mitigation actions.
    \item Interface with international networks for astrophysical transients, receiving, processing, and filtering incoming information, and triggering observations within four seconds of receiving an external alert.
    \item Provide preliminary data reduction and analysis, including full field-of-view (FoV) searches for transient and time-variable phenomena, along with quick-look data quality assessments.
    \item Automatically generate and revise the observational schedule during an observing night without human intervention, while keeping operators informed of the current schedule at all times.
    \item Enable fully automatic operation throughout the observing night in the absence of alarms.
    \item Support a full restart of the ACADA software within two minutes, restoring full monitoring and alarm functionality of all available AEs within five minutes.
    \item Designed to be operated by two individuals: one operator and one support astronomer, both located in a control room.
    \item Ensure that ACADA’s core functionality—required for system control and safe execution of observations—is available at least 99\% of the time during observation periods.
    \item Be deployable on any standard computing node running a Red Hat Linux derivative in the on-site data center (currently AlmaLinux 9.1\footnote{\url{https://almalinux.org}}) without requiring specialized hardware.
    \end{itemize}

From the requirements, interfaces, and CTAO top-level architecture, we developed a set of use cases that describe the interaction of users with the ACADA system. From our analysis, we identified four user roles for the system:
\begin{itemize}
    \item {\bf Operator:} Responsible for supervising and carrying out scheduled observations and calibrations during the night. This user troubleshoots problems, can modify the schedule, if necessary (e.g., weather, Science Alerts), and logs all activities.
    \item {\bf Support Astronomer:} Oversees and supports scheduling of observations from long-term to short-term, supervises reactions to external and internal science alerts, and checks science and data quality monitoring results.
    \item {\bf Configuration Manager:} Keeps track of the configuration of all instruments, part replacements, etc. Responsible for the central management of all documentation, logging, and configurations.
    \item {\bf Operator with Access Privileges (Operator+):} Has expert access to the instruments and software at the CTAO array sites and executes debugging and engineering activities in case of reported problems. This category also includes experts in the ACADA system.
    \item {\bf Engineer/Technician:} Like the Operator, but typically uses the system during daytime for Technical operations.
\end{itemize}

The use cases were first captured as UML use case diagrams and later specified as text-based use cases, including the main scenario description, alternative and exception paths, triggers, and pre- and post-conditions. More than 100 use cases were identified, including use cases describing system startup and shutdown, observation execution, and data acquisition and handling. A typical use case of ACADA can include scenarios from 10 to 20 steps. Use cases are used in ACADA as the guiding features for the continuous development and testing of ACADA at the product level (see Sec.~\ref{sec:devel-process}), around which we integrate developments of ACADA subsystems, and publish incremental versions of the ACADA product.

\subsection{Science Data Model}

The CTAO data model~\citep{cta:topleveldatamodel} defines the category of {\it Science Data}, which refers to data acquired, processed, or distributed to users. It describes nine data levels, denoted \textit{R0} and \textit{R1} or \textit{DLx}, where x is an integer indicating the processing level from 0 to 6. Each data level represents the degree of processing applied within the standard analysis workflow used to process observations.  \textit{R0} data are acquired by a controllable AE (telescope or other instrument) and remain in a device-specific form that has not yet been conditioned for transmission to ACADA. \textit{R1} data are raw data streams sent to ACADA, incorporating preliminary calibration and formatted into a standard wire format. \textit{DL0} data are standardized raw datasets stored by ACADA on disk, accompanied by the necessary metadata, and compressed into a file format suitable for long-term storage and subsequent retrieval. \textit{DL3} data are science-ready datasets generated by the DPPS for delivery to CTAO users. They contain selected extended air shower events, each with a single final set of reconstruction and discrimination parameters, together with associated Instrument Response Function (IRF) data describing the astronomical, environmental, and instrumental conditions required for scientific analysis. Further processing is carried out up to \textit{DL5} through SUSS capabilities. \textit{DL5} represents advanced scientific data products, including spectral energy distributions, light curves, sky maps, and phaseograms. The last data level, \textit{DL6}, corresponds to high-level products such as source catalogs and falls outside the scope of ACADA.

The SAG (see Sec. \ref{sec:SAG}) produces quick-look \textit{DL3} and \textit{DL5} data for use by the Support Astronomer and other CTAO stakeholders.

\section{ACADA System Architecture}
\label{architecture}

The design of the ACADA architecture, derived using MBSE techniques \citep{ACADA-Architecture}, addresses the ACADA requirements while facilitating the development of its functions in well-decoupled subsystems and components, each assigned to different members of the geographically distributed ACADA collaboration. ACADA is designed as a highly reliable and fault-tolerant system that implements the requirements described in Sec. \ref{requirements}.

\begin{figure}
	\centering 
	\includegraphics[width=0.4\textwidth, trim={0.5cm 1cm 2cm 0cm},clip]{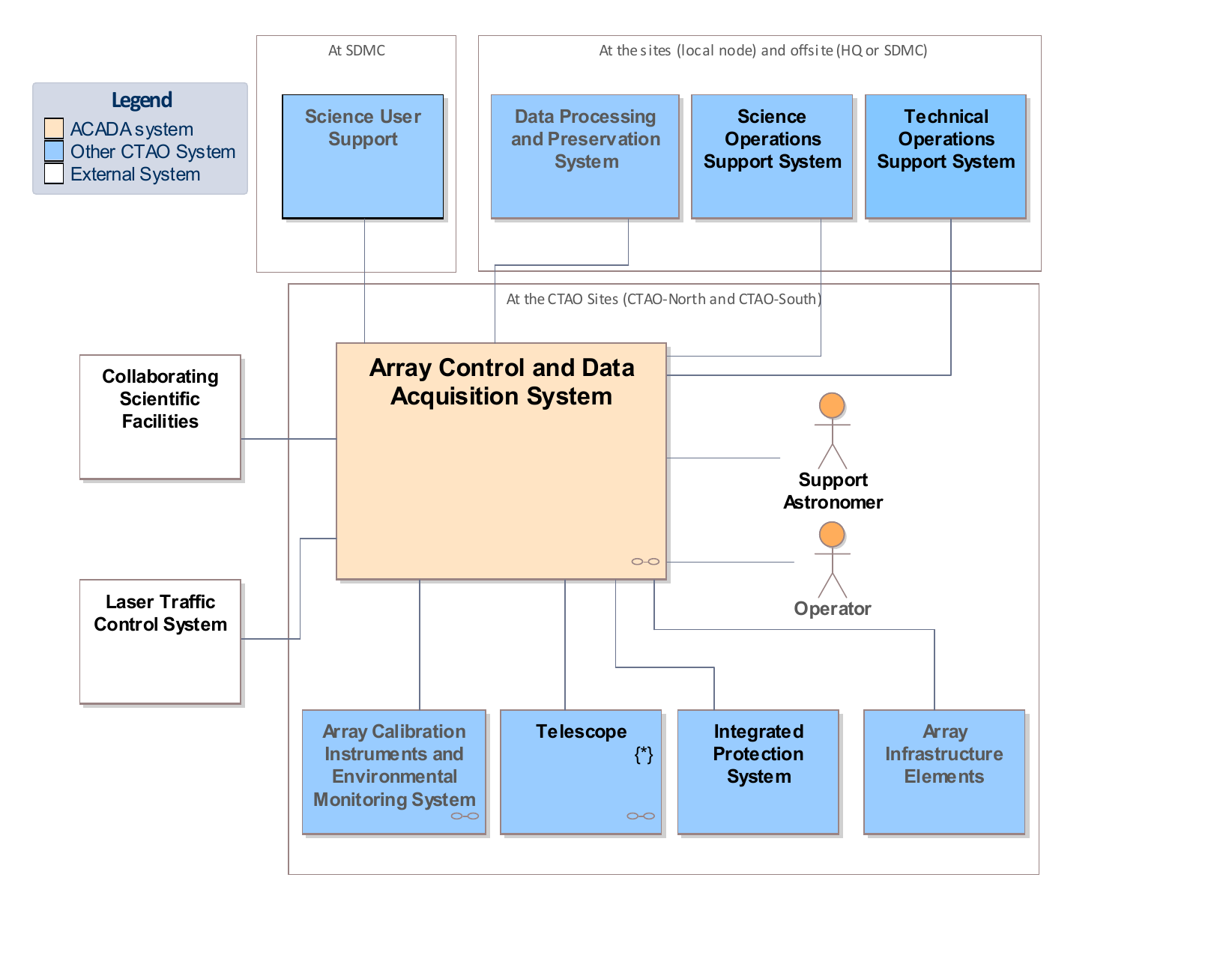}	
	\caption{ACADA system context. The systems external to ACADA are depicted in blue (when part of the CTAO System) and in white (when external to the CTAO). The stick figures represent the human actors interacting with the system.} 
	\label{fig:context}%
\end{figure}

ACADA is a key element at each CTAO site (CTAO-North and CTAO-South) and interfaces with numerous other CTAO systems. As illustrated in Fig.~\ref{fig:context}, ACADA interfaces with the following CTAO systems: the telescopes; the array calibration and environmental monitoring systems; the DPPS; the Science Operations Support System (SOSS); the SUSS; the Integrated Protection Systems (IPS); infrastructure elements such as the On-Site Data Center and power-management system; and the Technical Operations Support System (TOSS). A detailed description of the systems surrounding ACADA lies beyond the scope of this work; further information on the overall CTAO system architecture can be found in \citep{CTATopLevelArch}. In addition, ACADA directly interfaces with external collaborating scientific facilities (for handling external science alerts and coordinating schedules with other observatories) and with the laser traffic-control system at each CTAO site to coordinate laser operations (LIDARs and calibration devices).

ACADA is a distributed system built upon the Alma Common Software (ACS) framework \citep{ACS}, which serves as its core middleware. ACADA provides its functionality through components, each responsible for one or more clearly defined tasks. Most ACADA components are implemented as ACS components running on standard computing nodes within the on-site data center. The various ACADA components interact to deliver the functionality required by the system as a whole. ACADA relies on ACS for fundamental services such as component instantiation and inter-component communication. Each ACADA component has an associated supervisor that monitors its existence and integrity and orchestrates its replacement with a successor when necessary. Component supervision and replacement are organized hierarchically within a supervision-tree structure (see Sec.~\ref{rm}). The ACADA technology stack includes the following main elements:

\begin{itemize}
    \renewcommand*\thefootnote{\alph{footnote}}
    \item {\bf Middleware:} ACS\footnote{\url{https://confluence.alma.cl/display/ICTACS/}}, OPC UA\footnote{\url{https://opcfoundation.org}} (for monitoring-data access only)
    \item {\bf Programming languages:} Python, Java, C++, JavaScript (for the Human-Machine Interface (HMI) front-end only)
    \item {\bf Databases:} MySQL\footnote{\url{https://www.mysql.com}}, MongoDB\footnote{\url{https://www.mongodb.com}}, Redis\footnote{\url{https://redis.io}}, Cassandra\footnote{\url{https://cassandra.apache.org/_/index.html}}
    \item {\bf Messaging and serialization:} Apache Kafka\footnote{\url{https://kafka.apache.org}}, ZeroMQ\footnote{\url{https://zeromq.org}}, FITS\footnote{\url{https://fits.gsfc.nasa.gov/fits_home.html}}, Google Protocol Buffers\footnote{\url{https://protobuf.dev}}, REST, JSON, XML, and CORBA (via ACS)
    \item {\bf Workload management system:} Slurm\footnote{\url{https://slurm.schedmd.com}}
    \item {\bf Containers:} Docker\footnote{\url{https://www.docker.com}}
    \item {\bf Other:} ESO's Integrated Alarm System(IAS)\footnote{\url{https://integratedalarmsystem-group.github.io}}
\end{itemize}

\begin{figure*}
    \centering 
    \includegraphics[width=0.9\textwidth, trim={0cm 3cm 0cm 0cm},clip]{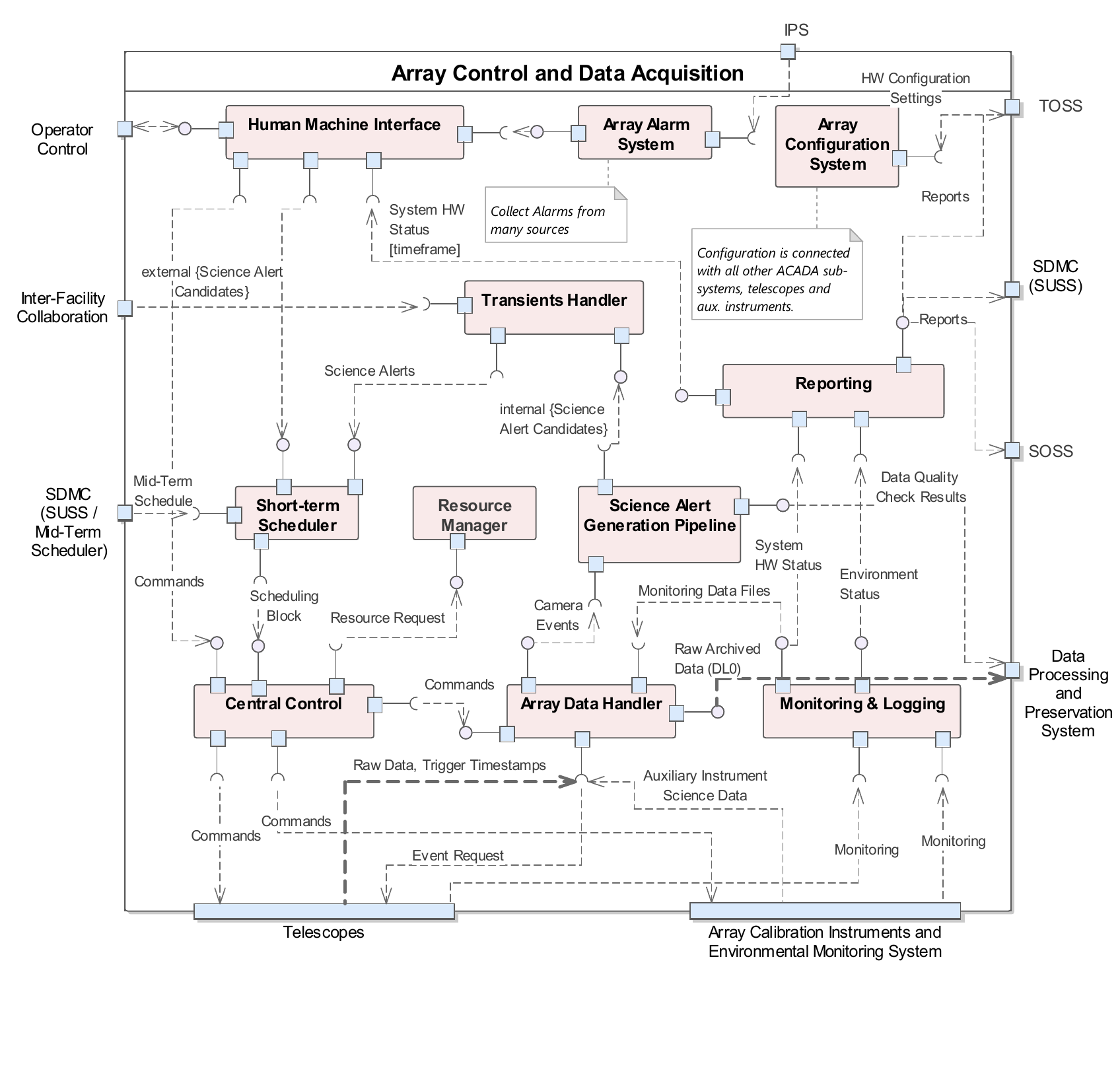}	
    \caption{Logical view of ACADA, representing the main components of the system. Only the highest-level components and the most relevant data elements and interfaces are shown. The diagram uses the UML component notation, and the basic entities are the software components, depicted as pink boxes. Blue squares are “ports” that identify the interfaces of the system. Those ports in the boundary of the ACADA identify the external interfaces of the system, and those in the components identify the internal interfaces. The dashed lines show the flow of data elements and the arrows the direction of the flow. The thicker data flow line indicates the flow of Cherenkov camera data through ACADA. The ACADA Command Line Interface and the ACADA Task Manager are not shown in this diagram.}
	\label{fig:components}%
\end{figure*}

The ACADA subsystems are illustrated in Fig.~\ref{fig:components} and described in the remainder of this section.

\subsection{The Resource Manager} 
\label{rm}

The Resource Manager (RM) is the subsystem responsible for supervising all other systems of ACADA, as well as any external systems under ACADA’s control, such as the control systems of the telescopes. Another key function of the RM is managing the allocation of telescopes to subarrays. The RM is implemented in Java. Its most important components are the following:

\begin{itemize}
    \item {\bf Array Elements Supervisor:} Provides the capability to monitor the software components of AEs. The Array Elements Supervisor tracks the availability of telescopes and auxiliary devices for service. If an AE has been taken out of service due to problems, maintenance, or other reasons, this information is recorded by the Array Elements Supervisor and can be used by the STS (see Sec.~\ref{sts}).
    \item {\bf Component Starter:} Provides services to any supervisor component that needs to instantiate and start another component. The Component Starter can identify an appropriate computing node to host a particular supervised component. It assists the supervisor component in getting the new component up and running, providing the necessary runtime initialization parameters, which may be component-specific.
    \item {\bf Computing Node Registry:} Tracks the availability and utilization of computing nodes used by ACADA. It is used by the operator HMI to obtain a list of all nodes utilized by ACADA components, check individual availability and idle status, and take computing nodes out of service or reactivate them.
    \item {\bf Persistence Service:} A service to persist the state of RM, Central Control (CC, see Sec.~\ref{sec:cc}), and components of other ACADA subsystems that require it. Some components, when replacing a pre-existing component with the same role, need to include the state of the predecessor to function properly; this is facilitated by the Persistence Service. The service is backed by a high-performance database using Redis technology.
    \item {\bf Role Lookup Service:} Service that allows the look-up of components by role name, facilitating the identification of a service component by functionality. Provides convenient access to the currently active member of the successor chain of an ACADA component. The Role Lookup Service also maintains references to all instances of all components acting in a particular role, as well as the specific control ticket each holds. Almost every ACADA component uses the role look-up functionality of the Resource Manager.
    \item {\bf Root Supervisor:} The root node of the ACADA supervision tree. All components started by the Root Supervisor are deployed to run continuously and are restarted by the supervisor in case of a problem. For fault tolerance, the Root Supervisor is instantiated in two instances (“cold” and “hot”), deployed on different computing nodes. The list of all components to be supervised by the Root Supervisor is statically configured through the ACS configuration database.
    \item {\bf Supervision Tree Support Libraries:} Used to monitor the status of system components, providing mechanisms to react to and recover from failed software components without affecting other ACADA subsystems. Supervision is organized as a tree structure, with the root node within the RM.
\end{itemize}

\begin{figure}
	\centering 
	\includegraphics[width=0.4\textwidth]{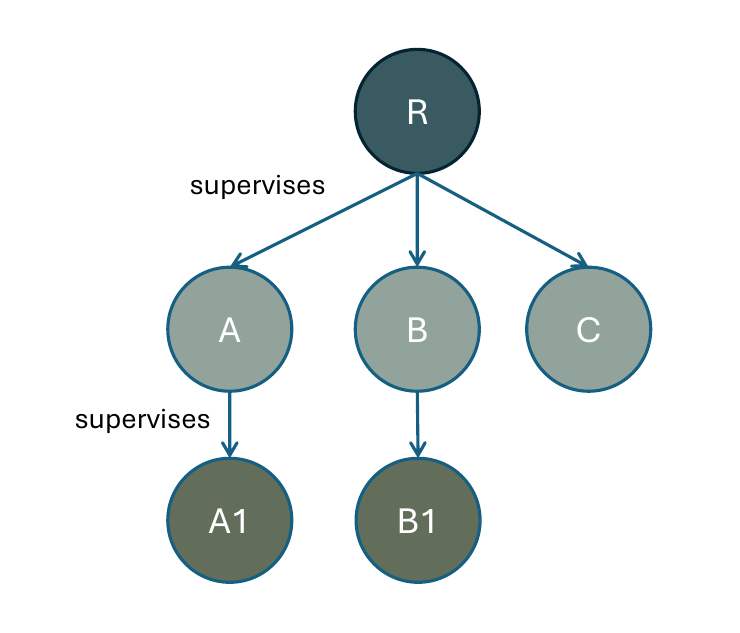}	
	\caption{Sketch of a supervision tree.} 
	\label{fig:supervision-tree}%
\end{figure}

ACADA uses supervision trees to handle potential failures and become fault-tolerant. The RM includes the root node of a supervision tree scheme (see Fig.~\ref{fig:supervision-tree}.) In a supervision tree, at run-time, the system is structured as a tree, with each tree node supervising its child nodes. In this context, to “supervise” means:
\begin{itemize}
    \item {\it Initialise}: Request the supervised component to start.
    \item {\it Poll} status: Periodically check the existence and status of the supervised component.
    \item {\it Replace} on problem: Replace the supervised component with a successor if it disappears or reaches an error state.
\end{itemize}

The components of the ACADA system base their instantiation strategy on the supervision tree. The first node in the tree is the Root Supervisor component described earlier. The Root Supervisor instantiates and supervises the elements under its control. Any of these second-layer components can instantiate and supervise additional elements. A supervision tree node can replace a supervised component with a successor component, with the capability to retrieve the relevant state data of the predecessor, supported by the Persistence Service.

The benefit of this approach is that it eliminates the risk of the entire system restarting in the event of a crash of an individual component and provides a mechanism for rapid system startup. This also implies that all ACADA components must utilize these services.

More details on RM are provided in \cite{ACADA-RMCC}.

\subsection{Central Control} 
\label{sec:cc}

The CC subsystem is responsible for coordinating control operations at the CTAO sites. 

Scheduling blocks (SBs) are the fundamental units for scheduling and executing observations with the CTAO. An SB is composed of one or more Observation Blocks (OBs).  OBs are the smallest scheduling units, representing a continuous observation with a subarray during which science data is collected on the target of the parent SB. CC orchestrates the observations by executing the SBs submitted by the STS. During the execution of an OB, the camera configuration and telescope target remain fixed. 

The CC uses the information contained in the SB data structures to send commands to the telescopes and other controllable AEs, and to monitor the execution of those operations. The CC also informs the Array Data Handler (ADH) and the SAG pipeline of upcoming observations so that they can prepare to handle and analyze data, respectively. 

The CC cooperates with the RM to request the resources that are required for the operations of a particular SB. Normally, the CC ensures that the SB finishes and then informs the Short-term Scheduler (STS, see Sec. \ref{sts}) about the outcome. However, the CC can receive cancel commands for an ongoing execution of SBs from the STS or from the operator.

The conditions for the continued execution of SBs are continuously monitored by the CC using the ACADA monitoring subsystem. If necessary, executions are canceled. The CC informs the HMI, the STS, and the SAG about the status of SB executions. In addition, the CC is responsible for: 
\begin{itemize}
\item Distributing the operations into simultaneously operating subarrays as requested by the scheduler.
\item Implementing any required operation modes (e.g., wobble mode, divergent pointing).
\item Informing the operator user interface about the status of the execution of the SBs.
\item Checking conditions that may require stopping a currently ongoing SB. These conditions can be related to the environmental factors, the loss of a functioning telescope, or more sophisticated reasons (e.g., energy threshold).
\item Providing a convenient scripting language to implement and execute observing modes as scripts.
\item Maintaining a script repository to store various operation scripts that define observation tasks to be performed.
\end{itemize}

The CC system is implemented using a combination of Java and Python programming languages. Java is used for most of the components except for the operation scripts and the script environment/interpreter, where Python is used.  

During the observation, the CC checks the real-time external factors that might affect the observation, for example, worsening weather conditions, and stops the observation if necessary. A plugin-based infrastructure is used to easily extend the conditions the CC needs to check during the CC execution.

More details on CC are provided in \cite{ACADA-RMCC}.

\subsection{Human-Machine Interface} 

The ACADA HMI provides a comprehensive view of the status of observations to the ACADA users (in particular, to operators and support astronomers, see Sec. \ref{requirements}) located in the control room of the CTAO installations, offering tools to supervise and interact with the AEs. This includes (but is not limited to) enabling the operator and support astronomer to:

\begin{itemize}
\item Monitor and control the life-cycle of ACS and all ACADA subsystems (e.g., start, stop, restart operations).
\item Manipulate the short-term schedule via STS.
\item Manually modify running observation blocks via CC.
\item Monitor the health of AEs and their subsystems, including controllable systems, software processes, environmental conditions, etc., via MON.
\item Change the operational state of AEs via RM. 
\item Monitor the scientific quality of ongoing observations via SAG.
\item Manually react to scientific alerts and manage priorities via Transients Handler (TH, see Sec.~\ref{sec:th}).
\item Diagnose potential operational problems via the Array Alarm System (AAS, see Sec. \ref{sec:AAS}).
\end{itemize}

The HMI subsystem includes two main building blocks: a \textbf{Data Manager (DM)}, which contains the code that interfaces with other ACADA subsystems, providing bidirectional communication and data aggregation; and a \textbf{Frontend Manager (FM)} block, which includes the functionality to display information to the operator and allow them to issue commands to the system. The FM provides interactive visual interfaces via a web browser, served by one or more web servers.

The DM comprises several services, which may run on separate CPUs. Each service represents a class of interfaces for different ACADA sub-systems. The exact number of services depends on the evolving complexity of interfaces to other subsystems. A conservative estimate is 10-20 services. The configuration is flexible and modular and may be modified as needed via a configuration file. The FM comprises several independent web servers, and it is easily scalable depending on the needs of the CTAO installations and the system configuration. An in-memory Redis database serves as a data buffer and communication broker between the DM and FM components. 

\begin{figure*}
	\centering 
	\includegraphics[width=0.9\textwidth]{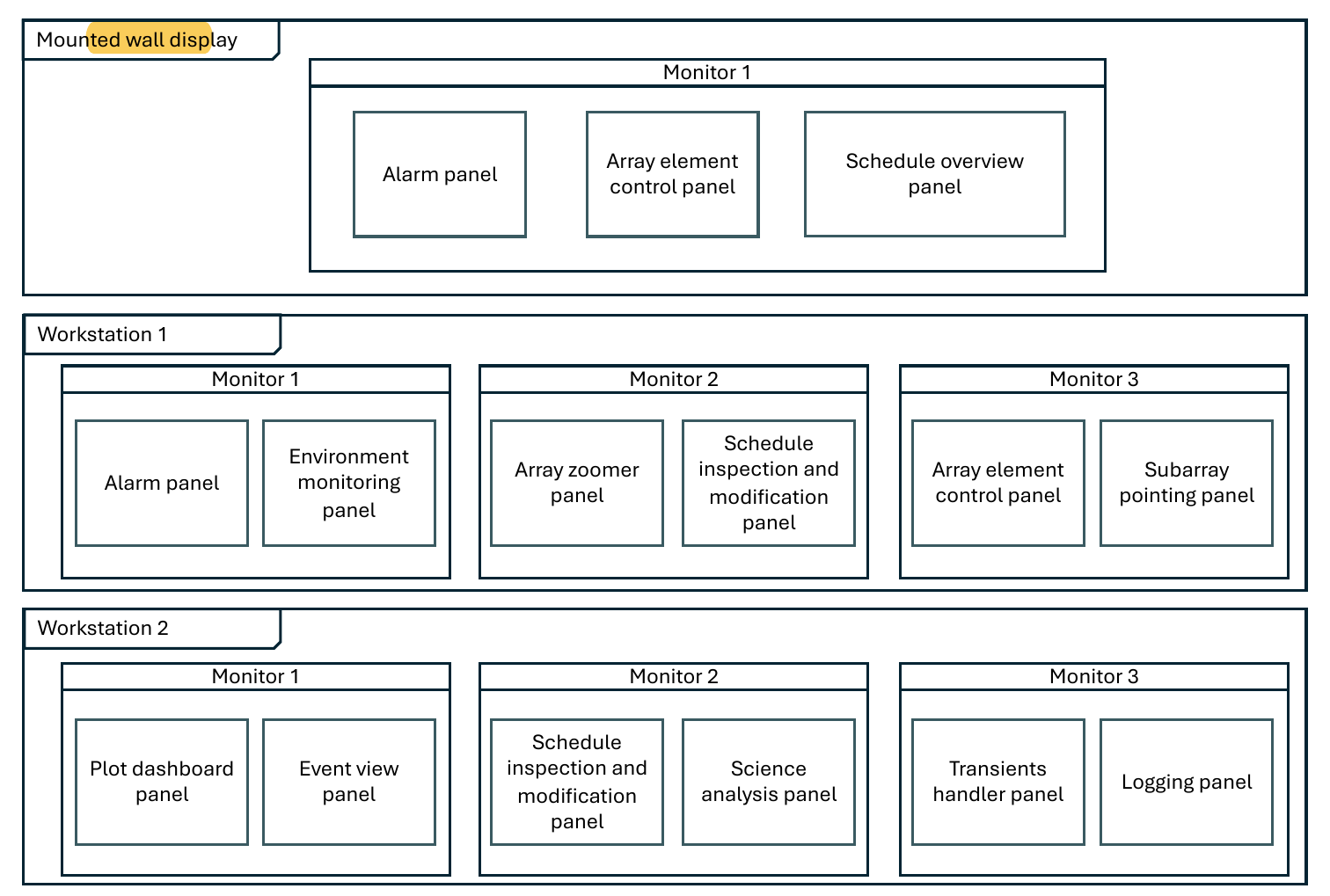}	
	\caption{Sketch distribution of the ACADA HMI panels, and they will be distributed in the workstation monitors and wall panels in the CTAO control rooms.}
	\label{fig:hmi-panels}%
\end{figure*}

\begin{figure*}
	\centering 
	\includegraphics[width=0.9\textwidth]{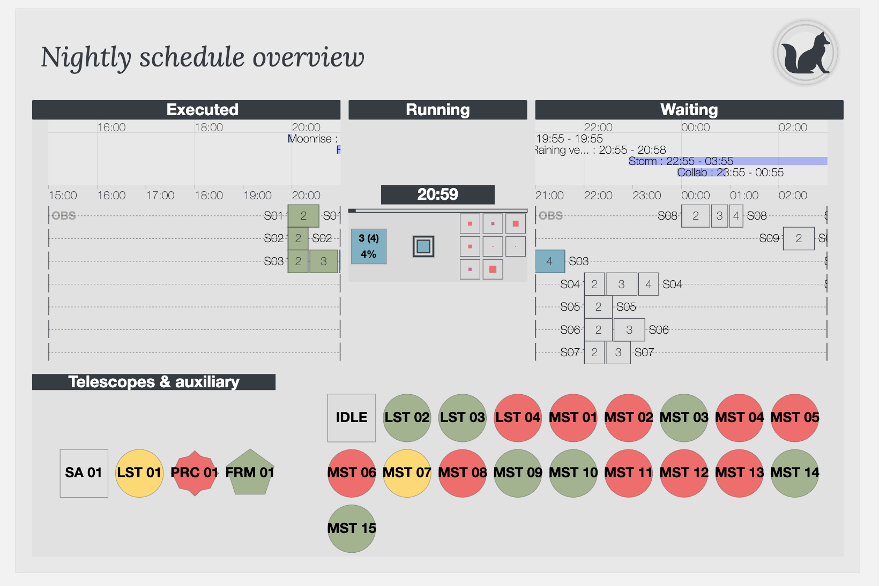}	
	\caption{The Schedule Overview Panel shows the planned SBs and OBs for the night. The main top area displays OBs as rectangles, grouped into SBs (as indicated by “S01” for the first SB). The right-hand area corresponds to blocks waiting to be executed; in the middle, the blocks currently being executed are shown; and on the left, those blocks that have finished execution are displayed. The colors represent the execution status of the blocks: Gray – waiting to be executed; Blue – currently executing; Green – finished successfully; Purple – canceled or unplanned (not shown in this example); Red – failed execution (not shown in this example). The middle section of blocks currently being executed also shows the execution phase of the SB. In this example, eight possible phases are shown, indicated by the small rectangles to the left and right of the blue observation block. The bottom part of the display shows the allocation of AEs to the blocks currently being executed, as well as those AEs not currently allocated to any running block. The different shapes represent different AE types, consistent with the visual language used in other panels.} 
	\label{fig:hmi-sched}%
\end{figure*}

The ACADA HMI design considers the deployment of 15 panels, displayed on two workstations (each with three monitors) and a mounted wall display (see Fig.~\ref{fig:hmi-panels}). A description of the content of all these panels is beyond the scope of this work; here, we describe the \textit{Schedule Overview Panel} as an example (Fig.~\ref{fig:hmi-sched}).

\subsection{Array Data Handler} 

The ADH subsystem is responsible for handling the stream of data from the array instrumentation. ADH includes components to handle the Cherenkov camera data, as well as to reduce the received data volume on the fly. Additionally, it provides array-level trigger capabilities via the software array trigger (SWAT), and hands over the data to the DPPS node located onsite, which eventually transfers this data offsite. The building blocks of ADH are the following: 
\begin{itemize}
\item The {\bf The Science Data Handler (SDH)} components receive raw Cherenkov data (R1) from the camera servers, zero-suppress them, forward them to SAG, and, in parallel, compress and write it to the on-site storage. It handles at least one stream per telescope.
\item The {\bf Auxiliary Instrumentation Data Handler} components receive data from devices that produce a data rate too high to be handled by the Monitoring System (MON, see Sec. \ref{mon}), which supports monitoring of data items at a rate up to a few Hz. The component formats the incoming data according to the DL0 data format and writes it to the on-site storage.  In addition, once a day, it creates a copy of the data in the MON database into the on-site repository, also in DL0 format. 
\item The {SWAT} reads Cherenkov events’ timestamps from the Cherenkov cameras. It seeks coincidences and provides feedback to the camera server to inform them whether an event is part of an array event or only local to that telescope. In normal operation, most local events are to be discarded while array events are to be transferred to the SDHs and written to the on-site repository. A fraction of local events may be accepted, e.g., for calibration purposes.
\item The {\bf Data Volume Reduction (DVR)} library is used by the SDH to convert R1 Cherenkov events to DL0. To reach the target data reduction of a factor greater than 15, which is required for the long-term full-scale operation of the CTAO, certain information must be removed during this conversion. Hence, the primary task of the DVR library is to identify image pixels and waveform parts that contain the information most relevant for downstream analysis. To achieve robust operation, the DVR library has to take into account camera-level and pixel-level health metrics \citep{dvr:2025}. 
\item The {\bf Supervisor Component} is the interface between ADH and CC. It is responsible for receiving commands from CC and executing them by controlling lower-level components. When a new subarray configuration (from a new SB) is received, it starts or connects to the relevant components and configures them. When a new observing block starts, it receives an order to move to a new dataset from CC. During normal operations, it monitors low-level components, takes action upon failures, and produces statistical summaries for CC.

\end{itemize}

The ADH system is implemented using a combination of Python and C++ programming languages. Python is used for the Supervisor and Auxiliary Instrumentation Data Handlers components, while C++ is used for the SDH and SWAT, as they require maximum performance. All ADH components are implemented as ACS Components, except from the Auxiliary Instrumentation Data Handler, which is a command-line tool.

\subsection{Transients Handler} 
\label{sec:th}

The Transients Handler (TH) subsystem is responsible for managing internal (generated by the SAG, see Sec.~\ref{sec:SAG}) and external (from outside ACADA) science alert candidates by filtering, processing, and ranking them, with the capability of requesting schedule updates, including immediate reaction if needed. To that end, the TH implements the following functions:

\begin{itemize}
    \item {\textbf{Broker System:}} The entry point for external and internal science alert candidates. It keeps a stable connection to other scientific facilities to receive and submit science alerts. Furthermore, it performs basic verification and validation steps on incoming alerts (e.g., de-duplication). The Broker System consists of a broadcaster and a receiver node. The broadcaster allows the scientific community to be informed about CTAO alerts. Incoming alert candidates are passed on for further processing within the TH.
    \item {\bf Processing of alerts:} Incoming alert candidates are processed in a set of pipelines. The alert candidates are matched to \textit{science configurations} (typically according to accepted observation proposals), which can specify simple criteria as well as complex processing instructions like correlations with catalogues. The science configurations are stored in the Science Configuration Database. A decision on follow-up decisions is taken. The priority of the science alert observation is determined by taking into account the ranking of observation proposals and Observatory policies.
    \item {\bf Observation request:} It triggers the creation of a dedicated SB for the science alert, and its insertion to request the STS to schedule observations of the science alert. These observations may indicate the need to immediately start observing (prompt science alerts). \textit{Note:} The STS can potentially refuse to schedule the new observations if the priorities of ongoing observations are higher, or if the observation conditions do not match. 
    \item {\bf Informing the Support Astronomer:} Inform the Support Astronomer when science alerts happen, including information to allow the support astronomer to evaluate the relevance of the alert.
\end{itemize}

Both external and internal science alert candidates are handled automatically by the TH. With external science alerts, the system receives the alert data from various communication channels and formats it into a standard format for transient alert evaluation. Regarding internal science alerts, in addition to the formatting functionality, the system also receives maps (positions of significance hotspots, full significance maps) generated by the SAG pipeline regularly, and sends the information to the transient evaluation component in the TH.

The TH provides an interface to the SUSS system to configure the TH (e.g., number of science alerts, which pipeline version to execute, switch on/off, etc.) and provides suggestions to the SUSS for follow-up observations beyond the night, along with night performance metrics (via the ACADA Reporting System, see Sec.~\ref{sec:rep}).

See \cite{ACADA-TH} for more details on the ACADA TH subsystem.

\subsection{Science Alert Generation Pipeline}
\label{sec:SAG}

SAG is a pipeline running online that performs a quick look data processing and analysis of the acquired data, and produces data quality indicators. A fundamental role of the SAG pipeline is to check data acquired by the CTAO on the fly and signal when excess counts are found from a gamma-ray source. This can serve both to generate internal candidate science alerts (alerts raised by the CTAO), and to provide fast feedback from external alerts. The SAG can inform the TH system about a science alert candidate, which, after the TH processing, can cause the observations to be rescheduled, and/or other observatories to be made aware of the event seen by the CTAO. 

SAG performs science monitoring (generation of sky maps and light curves) for the support astronomer located in the control room, and also generates data quality information, which can be displayed in the HMI and trigger data quality alarms in the AAS (see Sec. \ref{sec:AAS}). SAG has 20 seconds from when it receives the data to perform the analysis.

The SAG is composed of the following modules:

\begin{itemize}
    \item {\bf Image parameter extractor and low-level reconstruction pipeline (SAG-RECO)}. The SAG-RECO implements a pipeline for fast reconstruction to process individual DL0 events and produces the corresponding DL3 data with a latency of less than 15 s (the other 5 seconds is employed, in parallel by the other SAG pipelines). SAG-RECO provides a component to integrate DL0 data cubes (signal intensity and arrival times per pixel) into images, clean images, and then extract image parameters.
    The parameters and images are then delivered to SAG-DQ. The high-level parameters, including the energy, direction, and a gamma-ray-like classification score, are computed using machine learning algorithms (Random Forest) trained on Monte Carlo simulations. The final step consists of performing an event selection in order to ensure quality and select gamma-like events, and provide those events to SAG-SCI. This component can extract the image parameters at a rate higher than 1000 events/s/CPU/telescope. See \cite{ACADA-SAG-reco} for more details.
    \item {\bf Data quality pipeline (SAG-DQ)} The SAG system also performs an online data quality analysis to assess the quality of the data during the data acquisition. It allows the user to define, through XML configuration files, the structure of the input data and, for each input data field, which data quality results should be produced. There are two types of data quality results: statistical data display (such as distributions of parameters) and applied data quality checks. These checks verify whether the input data values satisfy predefined constraints. For example, if the pixel value of a telescope’s camera exceeds an upper bound threshold, a warning or alarm will be generated.
    \item {\bf High-level analysis/scientific pipelines (SAG-SCI)}. The scientific pipelines (SAG-SCI) are based on a framework designed for the development of real-time scientific analysis pipelines~\citep{PARMIGGIANI2022100570}. Using this framework, the integration of existing science tools is simplified, providing a common pipeline architecture and automatisms. The framework can be configured with new or existing science tools implemented in different programming languages. The scientific analyses are performed in parallel and can be prioritized. The pipelines distinguish between High-Level Analysis pipelines for Science Monitoring to produce science quick look (DL4 and DL5 data levels, generation of sky maps and light curves) information for the Support Astronomer, and SAG pipelines to produce science alert candidates. A \textit{Workload Manager} manages the computing workload of all the pipelines, allowing them to scale on a cluster of machines.
    \item{\bf Supervisor (SAG-SUP)} The SAG-SUP is the interface with the rest of the ACADA system. It supervises the operation of the SAG pipeline component associated with a subarray, and manages the generation of scientific monitoring results for the HMI. It provides input to the ACADA Reporting Subsystems for report generation for the DPPS and SUSS. SAG-SUP is also responsible for computing and updating the Good Time Interval (GTI) for each subarray. The GTI represents the periods in which the array meets all required conditions for valid gamma-ray data acquisition. These conditions include telescope tracking status, nominal environmental parameters (e.g., night sky background, cloudiness, humidity), and acceptable DL1/DL2 data quality. The GTI is used to filter valid events, supporting the generation of reliable science alerts.
\end{itemize}

SAG uses Gammapy \citep{gammapy:2023} in the SAG-SCI pipeline. \cite{ACADA-SAG} provides more details on this ACADA subsystem.

\subsection{Short-Term Scheduler} 
\label{sts}

The STS subsystem is responsible for deciding, in a short time, how to group and use the telescopes of a CTAO installation to perform nightly operations. 

Within the STS, the \textbf{Short-term Planner} component is responsible for making real-time decisions on how to group and use the CTAO telescopes to perform nightly operations, according to a previously supplied Mid-Term Schedule (MTS). This component is also responsible for real-time reactions to environmental changes and for processing external or internal science alerts provided by the Transients Handler. The short-term planner generates OBs and appends them to the SBs supplied by the MTS. The STS acts as a queuing system and sends SBs to the CC subsystem when their execution time arrives. The STARS (Scheduling Telescopes as Autonomous Robotic System)\footnote{https://stars.ieec.cat/home} library is used in the Short-term Planner for scheduling optimization. STARS is a software project designed to help observatories plan astronomical observations and modify them, regardless of the telescope type and the constraints involved. The STARS \textit{Core} library utilizes artificial intelligence concepts, providing the short-term planner functions to calculate the observable periods of SBs based on their constraints and to schedule them while optimizing for the defined objectives.

The CTAO will operate several lasers as part of the environmental monitoring, atmospheric characterization, and array-level calibration operations. Other instruments at the ORM and Paranal sites also operate lasers. This requires coordination between CTAO's intended observations and the neighboring telescopes. To that end, the \textbf{Laser Coordination Handler} component negotiates, in a short time, with each site's laser traffic control system the intended list of shots for lasers external to the CTAO and the plans for CTAO’s own laser operations.

Another component, the \textbf{Scientific Installations Coordination Handler}, is designed for synchronizing CTAO-North and CTAO-South observations, for importing observation plans from other scientific facilities, as well as for exporting observation plans from the CTAO to other observatories.

\subsection{Monitoring and Logging Systems}
\label{mon}

The Monitoring and Logging Systems (collectively referred to as MON) are components providing services for monitoring and logging AE and ACADA internal data. The monitoring components acquire data items from the telescopes and other devices deployed at the CTAO array sites and make these data immediately available to the HMI and the SAG for quality checks.

The \textbf{Monitoring Storage} uses a solution based on Apache Kafka for system integration and a data processing approach based on an open architecture with demonstrated scalability and reliability characteristics following modern Industry 4.0 standards. A Cassandra database is used for mid-term (up to 6 months) local storage of time-series data. Long-term storage is taken care of by the DPPS system and is outside the scope of this work.

The \textbf{Monitoring System} is also responsible for long-term storage of the execution history (states) of SBs and of OBs, as well as Observation Block phase histories. Such data points also serve as redundancies for the HMI, e.g., in cases of system restart and recovery.

Finally, \textbf{Logging System} includes capabilities for acquiring and storing logging data of AEs and ACADA, and a log analyzer to help identify issues and support preventive maintenance activities. 

See~\cite{ACADA-MON} for more details.

\subsection{Array Alarm System}
\label{sec:AAS}

AAS is a subsystem that provides the service that gathers, filters, exposes, and persists all the relevant alarms raised by both the ACADA processes and AEs under the supervision of the ACADA system. This subsystem shares design elements with the MON subsystem and, in particular, uses Apache Kafka and Cassandra as a message broker and database, respectively.

The system includes the following components:

\begin{itemize}
    \item {\bf Alarm Collector:} Gathers ACS or OPC-UA alarms raised by AEs or ACADA components. The design of the Alarm Collector encompasses features like shelving, unshelving, and acknowledgment, providing operators the ability to interact with the status of each alarm. Moreover, it incorporates functions such as Out-of-Service and Return-to-Service. The alarm Collector can send requests to the CC to execute operational scripts to trigger automatic alarm mitigation operations.
    \item {\bf Alarm Filter:}  The Alarm Filter provides mechanisms to create alarms. By combining multiple monitoring values and/or alarm events, the Alarm Filter is capable of triggering complex alarms according to a set of defined rules. The Alarm Filter makes use of the Alarm Rules Database, a database defining the alarm reduction rules for the Alarm Filter. The implementation of the Alarm Filter component utilizes the IAS~\citep{IAS}.  
    \item{ \bf Alarm Storage:}	The purpose of the Alarm Storage is to obtain alarm data from the Kafka topics supplied by the Alarm Collector and conduct the insertion query into the Cassandra database. Provides a local repository for storing alarms for up to six months. Permanent storage of alarms is taken care of outside ACADA by the DPPS.
\end{itemize}

The AAS is capable of identifying new alarms based on the analysis and correlation of the system software logs acquired by the MON Logging System (see Sec.~\ref{mon}). Alarm information can be visualized by the operator via the HMI. 

See~\cite{ACADA-MON} for more details.

\subsection{Array Configuration System} 
\label{sec:cdb}

The Array Configuration System, also known as the ACADA configuration database (CDB), stores and distributes the configuration datasets of the ACADA system components and the supervised AE systems. The datasets include the software deployment configuration and the instrumentation settings. 

The CDB is implemented as an OpenAPI\footnote{\url{https://www.openapis.org/}} compatible light-weight service with the SQL backend and FastAPI\footnote{\url{https://fastapi.tiangolo.com/}} web server connecting to the database using the SQLAlchemy\footnote{\url{https://www.sqlalchemy.org/}} object-relational mapping library.

Configuration data are primarily stored as JSON strings validated by the JSON schema\footnote{\url{https://json-schema.org/}}; the system also supports XML strings for existing legacy datasets.

CDB supports configuration versioning, validation, and transactions, allowing the integrity of the configuration set(s) to be maintained continuously. The rich tooling and plug-in architecture allows external users to utilize different backends and interfaces (e.g., command line interface) during the design and development of configuration data models and handling. The auxiliary tools provide CDB clients for CTAO-approved programming languages, as well as libraries for cross-language data modeling and code generation using Pydantic\footnote{\url{https://pydantic.dev/}} and JSON, which include agentic assistants based on large language models~\citep{Kostunin:2025rlk}.

\subsection{Reporting System} 
\label{sec:rep}

The Reporting System (REP) is responsible for gathering the relevant data from the other ACADA subsystems to produce status and quality reports of the ACADA operations, both for the HMI and for external systems. The implementation of this subsystem is planned for future versions of ACADA.

It will include the following building blocks:

\begin{itemize}
    \item {\bf ShiftLog Tool (Logbook):} A logbook that permits users to record events during a shift. Its capabilities will include recording events such as general observations, failures, weather downtime events, tests, and nightly summaries. It will include the database, user interface, and API for logbook functionality.
    \item {\bf Report Builder: } A component that receives data from other ACADA subsystems, stores and, where applicable, processes the data.
    \item {\bf Report Distributor:} The component that retrieves report content from the Report Builder component at defined intervals, to format the data as required by the internal and external interface(s), and deliver the reports to the respective internal and external subsystems.
\end{itemize}

REP will generate several reports covering many relevant performance indicators of the Observations of the night. The following non-exhaustive list names some of the indicators that will be included in the reports:

\begin{itemize}
    \item Numbers of available telescopes (total and per type) for science operations during the night, operational downtime, time spent per observation category (science observations, calibration, maintenance, etc.). Time spent not observing (e.g., switching observations).  
    \item Environmental and atmospheric status.  
    \item SAG pipeline performance indicators: analysis and data quality.
    \item Transients handling Indicators: Number of all external and internal science alerts received/accepted/observed.
    \item Data indicators: Number of DL0 files collected, total data volume of DL0 files collected, and integrity-check results (e.g., all DL0 files OK: yes/no).
    \item List of Shift Log events from the night.
\end{itemize}

\subsection{Other system elements}

The ACADA application design includes additional modules that complete the overall architecture and fulfill ACADA requirements. These are libraries and applications that support the operation of ACADA subsystems and provide a fine-grained interface with the system to support testing, commissioning, and troubleshooting. Most prominently, ACADA incorporates the following:

\begin{itemize}
    \item \textbf{Command-line Interface CLI}: Enables interaction with the ACADA subsystems.
    \item \textbf{Task Manager:} A component responsible for triggering and executing ACADA-level tasks, such as transferring acquired data to the DPPS, at certain times every day. 
\end{itemize}

\section{Development Process}
\label{sec:devel-process}

The ACADA development team consists of personnel from the CTAO Central Organisation and in-kind contributors (IKCs). The Central Organisation personnel are responsible for management and coordination, while the IKC teams deliver subsystems and supporting roles, such as the coordination of testing and integration activities. 

The ACADA product is delivered through a set of eight major releases with progressively enhanced capabilities, where intermediate minor versions will be made available at an approximate rate of one per month. The overall development process is incremental and iterative, with each major release typically completed within a six-month to two-year timeframe. The ACADA team works under a well-defined software quality assurance (SQA) process (see Sec.~\ref{sec:QA}). 
Within each major release, the team employs an \textit{agile} methodology with time-boxed sprints of one month with a defined list of features (normally use cases) to be delivered, tested, and integrated into the ACADA master branch (see Secs.~\ref{sec:devel} and ~\ref{sec:cont-testing}). At the end of each sprint, a review is carried out, often including a demonstration of ACADA capabilities. After that, ACADA is integrated and tested in the production environment (described in Sec.\ref{sec:integration}).

\subsection{Quality Assurance}
\label{sec:QA}

The quality assurance (QA) framework for CTAO Computing products, including ACADA, is crucial for maximizing the efficiency of the Observatory operations; therefore, an appropriate SQA activity is essential to maximize operational efficiency and ensure a high-quality, maintainable system. The ACADA SQA process consists of six stages that are embedded within its overall, iterative development lifecycle:

\begin{enumerate}
\item Definition of QA objectives for the upcoming ACADA release.
\item Internal review of the ACADA architecture.
\item Internal review of subsystem design updates.
\item Evaluation of static code analysis metrics and results.
\item System-level integration and verification of ACADA.
\item Quality-in-use evaluation. 
\end{enumerate}

The definition of quality objectives and specifications evolves with each ACADA release. This includes introducing new quality criteria and metrics, as well as refining target values for existing ones. The activity takes place during the Requirement Analysis phase of the ACADA development lifecycle. The goal is to achieve full coverage of all quality metrics by the final planned release, enabling an early focus on functionality while progressively enhancing code quality.

The internal architecture and design review examines and updates the ACADA architecture and subsystem designs. This review addresses issues from the previous release, accommodates requirement changes, and incorporates new details as they emerge. Upon completion of the design review, the project moves into the \textit{Implementation and Testing} phase.

During development, subsystem developers receive rapid, automated feedback on code quality through static analysis with SonarQube\footnote{\url{https://www.sonarsource.com/products/sonarqube/}}, using a customized quality gate. While continuous compliance with the gate is not mandatory, full compliance is required before final integration. In addition, subsystem-level quality metrics must meet their respective targets prior to release integration.

System-level metrics are assessed after complete system integration and verification of each ACADA release. Verification heavily relies on automated use-case tests, each traceably linked to its corresponding requirements. Quality-in-use evaluations of ACADA are planned for later stages of the Observatory’s construction phase.

For ACADA development, a dedicated Continuous Integration (CI) infrastructure, based on Jenkins\footnote{\url{https://www.jenkins.io/}}, has been established. This environment enables automated build, integration, and testing through dedicated pipelines, providing immediate feedback on the quality of submitted code (see Sec. \ref{sec:cont-testing}). However, since the CI setup currently operates on a single machine, it cannot fully reproduce ACADA’s distributed architecture. To overcome this limitation, a dedicated ACADA Test Cluster has been deployed, allowing testing and debugging in a distributed, production-like environment and enabling higher-level validation of the integrated software. Additionally, a Software Integration and Test Cluster supports off-site testing of integrated CTAO software packages together with AEs, extending testing beyond the ACADA scope.

This multi-layered quality assurance approach ensures that each ACADA release is functionally complete, validated, and ready for integration with broader Observatory systems, particularly CTAO AEs.

For every major ACADA release, progressively more stringent QA metrics are introduced, along with an expanded testing program involving AEs—including simulators, laboratory test stands, factory instrumentation, and deployed AE systems at the CTAO sites.

\subsection{Software Development}
\label{sec:devel}

While the core of ACADA is composed of multiple subsystems developed in independent repositories, the top-level ACADA repository contains integration components, including glue code, installation scripts, user manuals, system-level tests, and the ACADA CLI.

Subsystem versioning is organized via Git submodules, ensuring that each ACADA version is uniquely defined by the state of the main Git repository. Consequently, there is no need to track commit IDs or version numbers outside the repository. Each released version of ACADA corresponds to a specific commit tagged with a Git release tag.

\begin{figure*}
	\centering 
	\includegraphics[width=0.9\textwidth]{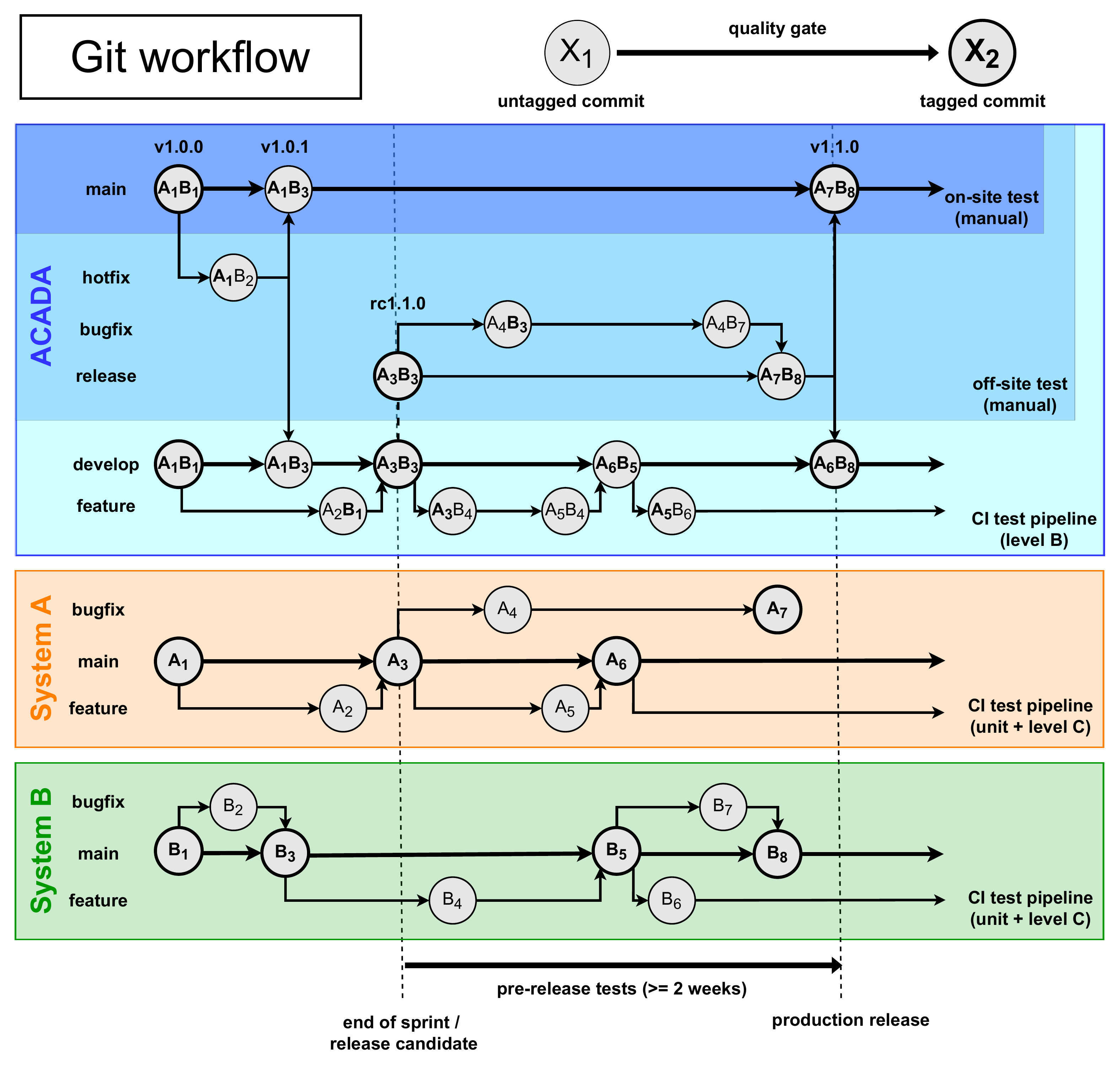}	
	\caption{An overview of the ACADA Git workflow.} 
	\label{fig:git_workflow}%
\end{figure*}

Developers from IKC teams working on subsystems are encouraged to follow the feature branch workflow, in which the default branch (main/master) is protected, and new features are introduced through merge requests (MRs). Before merging an MR, it must pass all automated validation steps—including SonarQube Quality Gates—and receive a positive human code review (see Sec. \ref{sec:QA}).

The main ACADA repository follows the Gitflow workflow, featuring two protected branches:

\begin{itemize}
    \item the \textbf{main branch}, which tracks official ACADA releases, and
    \item the \textbf{develop branch}, which tracks active development.
\end{itemize}

Feature branches are created from the develop branch and merged back upon completion. During development, submodules may temporarily reference any commit on a feature branch. However, before merging into develop, all submodules must reference tagged commits from their main branches. These tags serve as version anchors, supporting efficient rollback and improving traceability during retrospective code reviews by clearly linking subsystem changes.

Before a new ACADA release, comprehensive automatic and manual testing is conducted to validate functionality and stability. A release candidate branch is created from develop, allowing pre-release testing while restricting changes to critical bug fixes only. Merging rules for the release branch mirror those of the develop branch, and any changes applied to the release branch must also be merged back into develop to maintain consistency.

After the testing phase, the release candidate is merged into main, and a new ACADA version is formally released. Released versions are expected to remain stable; however, critical production issues trigger the creation of a hotfix branch derived from the release commit.

Subsystems providing hotfixes are permitted to reference tagged versions not yet merged into their main branches (e.g., if the subsystem’s main branch has progressed beyond the ACADA release version).
Once resolved, the hotfix branch is merged back into both main and develop, triggering a minor version release and maintaining alignment across all branches.

An overview of the workflow is shown in Fig.~\ref{fig:git_workflow}.

\subsection {Continuous testing}
\label{sec:cont-testing}

To maintain high software quality and prevent functional regressions, all ACADA subsystems, as well as the top-level ACADA repository, are subjected to continuous automated testing prior to merging any code changes.

Because test suites can require significant execution time, potentially disrupting the development workflow, tests are categorized by expected runtime into up to three groups, as described in Tab.~\ref{tab:testing}.
Moving tests further down the table allows faster iteration during development, but increases the risk of detecting faults at a later stage.

\begin{table}[ht]
\caption{ACADA Test categories.} 
\label{tab:testing}
\begin{center}    
\begin{tabular}{l l l}
\hline
\textbf{Name} & \textbf{Runtime}	& \textbf{When Executed}\\
\hline
Quick &	\textless~a few minutes	& Every push\\
\hline
Nightly	& \textless~a few hours	& Every night\\
\hline
Long &	\textless~a few days & 	Before a release\\
\hline
\end{tabular}
\end{center}
\end{table}

This tiered testing strategy ensures that rapid validation occurs with each code submission, while deeper, system-wide tests are periodically executed to guarantee overall reliability and readiness for release.

\subsection{Integration of ACADA with Array Elements}
\label{sec:integration}

The two CTAO arrays will comprise IACTs, along with a range of calibration and atmospheric characterization instruments. Within the IACT class, three telescope types are included: LSTs, MSTs, and SSTs. The calibration and atmospheric devices include illuminators, weather stations, Raman LIDARs, photometers, and all-sky cameras, among others. Integration between these AEs and ACADA will occur progressively, following the staged construction of the Observatory.

Once ACADA has been prepared and tested in the off-site test cluster and the deployment procedures have been validated, the integrated software can be deployed at the designated target site.
Target sites may include Observatory array locations, or laboratory facilities at partner institutes or companies developing CTAO AEs or subsystems. Throughout this process, all software packages and configuration parameters remain under strict configuration control, and deployment procedures are fully documented to ensure that installations are reproducible and verifiable.

Upon deployment, initial interface tests are conducted to verify the specific interactions between ACADA and the integrated AE. These tests focus primarily on subsystem monitoring, command transmission and status feedback, logging and alarm notification, data acquisition, and time synchronization between both systems.

For interface testing, only the components required to exercise the relevant interfaces are started, in order to avoid interference from unrelated subsystems that could obscure test results.
The goal is to confirm that all defined interfaces have been correctly implemented and are functioning as expected on both sides.

In addition to the interface tests, a second test suite focuses on system-level integration, involving multiple or all system components.
These tests culminate in an end-to-end validation of the available system capabilities.

This integration and testing methodology has already been successfully applied to the first CTAO prototype IACT, the first LST (LST-1), located at CTAO-North~\citep{ArrayElementIntLopez}. The approach is expected to mitigate architectural and design risks early, validate interface consistency, and provide prompt feedback on ACADA’s functionality and usability.

\section{Status}
\label{sec:status}

The ACADA team was established in February 2020. During the initial months, team members focused on the production of management, design, and quality assurance documentation. The ACADA design and development plans underwent a PDR in June 2020. The review panel determined the PDR to be a success and recommended that the ACADA team proceed to the detailed design phase.

The first official ACADA release (REL0) was announced in 2021, focusing on assembling the ACADA system for the first time by integrating existing prototype versions of the subsystems and incorporating them into the CI pipelines. The first use-case-based end-to-end tests were also implemented and executed, with the primary goal of exercising the ACADA software development lifecycle (SDLC) as a unified team for the first time. After REL0 was completed, a retrospective workshop was held to collect feedback from the extended ACADA team, followed by adjustments to the SDLC to better align with the team’s workflow and preferences. REL0 was deployed exclusively within the ACADA test environment.

ACADA Release 1 (REL1) was published in July 2023 and focused on functional requirements and single-telescope operations. It was designed to be fully capable of operating an individual CTAO LST. With the completion of REL1, approximately 50\% of the total ACADA codebase had been delivered, covering all core functionalities. In terms of use cases, those enabling the core operations of ACADA were already implemented, while many alternative and exception paths were not yet verified, and non-core use cases were deferred to future releases. Regarding formal requirement verification, around 25\% of ACADA requirements were verified at this stage. Since REL1 primarily aimed to deliver the core functionality, most non-functional requirements—particularly performance and scalability aspects—were scheduled for subsequent releases.

Following the release of REL1, an I\&T campaign was conducted by the ACADA and LST-1 telescope teams during fall 2023. During this campaign, ACADA and its telescope interfaces were tested extensively under realistic conditions, with the main objective of retiring design-related risks and validating interface implementations. Although the testing campaign achieved its main goals, several performance and interface issues were identified. Most internal ACADA issues were resolved through bug-fix releases during or shortly after the campaign. Performance optimizations, stability improvements, and minor design updates were implemented as part of the intermediate release, ACADA REL1.5, announced in December 2024.

The ACADA design, plans, and overall progress since the PDR were subsequently reviewed in a CDR held in January 2024. The review panel’s assessment was highly favorable, and all recommendations were addressed promptly, resulting in ACADA obtaining a formal “CDR Pass and Close-out” declaration by September 2024.

The current focus of the ACADA team is the realization of ACADA Release 2 (REL2), which aims to enable multi-telescope and automated operations support. Together with Release 3 (REL3), these versions will significantly enhance scalability, performance, and operational autonomy. Specifically, they will introduce the following new capabilities:

\begin{itemize}
    \item {\bf Multi-telescope support:} In REL2, ACADA will support the operation, data acquisition, and analysis of data from up to four telescopes of the same type. In REL3, this will be extended to five telescopes distributed across up to two subarrays. All three CTAO telescope types (LST, MST, SST) will be supported.
    \item {\bf Integration of other AEs:} ACADA will support stellar photometers and illuminators starting with REL2, and LIDAR systems in REL3.
    \item {\bf Automatic operations:} Beginning in REL2 and further enhanced in REL3, ACADA will be capable of executing nighttime observations autonomously, without human intervention, leveraging the capabilities of the STS. An expanding range of science cases will also be supported by the TH, which will coordinate with the STS to execute real-time observations.
    \item {\bf HMI panels:} Additional Human–Machine Interface (HMI) panels will be deployed, including dashboards for monitoring and assessing the health and status of AEs, with significant improvements in usability.
    \item {\bf Performance:} Restart times, real-time reaction to schedule changes, and other timing-related performance requirements will be optimized to achieve response times on the order of a few minutes.
    \item {\bf Availability:} The ACADA supervision tree will be fully implemented, enabling rapid detection and recovery of failing components and processes, thereby improving system robustness.
    \item {\bf Technical debt:} Identified issues and interface improvements from the integration and testing campaigns will be addressed, ensuring better performance, maintainability, and long-term stability.
\end{itemize}

Later ACADA releases (Release 4 and beyond) will expand on these developments by supporting larger telescope configurations, additional AEs, and more concurrent subarrays.
They will also introduce advanced alarm handling mechanisms and automatic generation of nightly Key Performance Indicators (KPIs) to further streamline Observatory operations.

\section{Summary and conclusions}
\label{summary}

After years of design and prototyping, the ACADA team has released software versions capable of operating a real CTAO telescope. To date, approximately 50\% of the ACADA codebase has been implemented, with a substantial fraction of use cases and requirements verified, and over 60\% of the code covered by automated tests. Successful integration tests with the LST-1 telescope and a favorable outcome of the CDR confirm the maturity and robustness of the software, while providing valuable feedback for further optimization.

Future ACADA releases will incrementally expand capabilities, enabling the operation of increasingly complex CTAO configurations and ensuring alignment with the CTAO construction milestones as the first telescopes are accepted into operations over the next few years.

\section*{Acknowledgements}
 
The authors gratefully acknowledge the support and contributions of all ACADA team members, as well as the members of the CTAO Computing, Systems Engineering, and Project Management departments, for their roles in establishing the ACADA management structure.
The authors also acknowledge the fundamental commitment to ACADA, through the provision of personnel and equipment, by Deutsches Elektronen-Synchrotron (DESY), Institut de Ciències de l’Espai (ICE/CSIC), Istituto Nazionale di Astrofisica (INAF), the University of Geneva – Department of Astronomy, Laboratoire d’Annecy de Physique des Particules (LAPP), Nicolaus Copernicus Astronomical Center of the Polish Academy of Sciences (CAMK), Max-Planck-Institut für Kernphysik (MPIK), the University of Potsdam, and Centro de Investigaciones Energéticas, Medioambientales y Tecnológicas (CIEMAT).

The authors further acknowledge the excellent support from the LST Collaboration, which enabled the ACADA team to conduct multiple tests of ACADA with the LST-1.
They also express gratitude to the team responsible for the computing cluster located in La Palma, owned by the University of Tokyo, where ACADA Release 1 (REL1) was deployed for testing with the LST-1.
Access to both the LST-1 and the computing cluster was fundamental to the successful execution of the ACADA test program. 

The authors gratefully acknowledge financial support from the agencies and organizations listed here: \url
{https://www.ctao.org/for-scientists/library/acknowledgments/}  




\bibliographystyle{elsarticle-harv} 
\bibliography{references}






\end{document}